\documentclass[preprint]{aastex6}
\usepackage{longtable}
\usepackage{epsfig}
\usepackage{graphicx}
\shorttitle{HR 2562B}
\shortauthors{Konopacky et al.}

\begin{document}

\title{Discovery of a Substellar Companion to the Nearby Debris Disk Host HR 2562}

\author{Quinn M. Konopacky\altaffilmark{1}, Julien
  Rameau\altaffilmark{2}, Gaspard Duch\^{e}ne\altaffilmark{3,4}, Joseph
  C. Filippazzo\altaffilmark{5}, Paige A. Giorla Godfrey\altaffilmark{6,7,8}, Christian Marois\altaffilmark{9,10}, Eric L. Nielsen\altaffilmark{11,12}, Laurent Pueyo\altaffilmark{5},
  Roman R. Rafikov\altaffilmark{13}, Emily L. Rice\altaffilmark{6,7,8}, Jason J. Wang\altaffilmark{3}, S. Mark Ammons\altaffilmark{14}, Vanessa P. Bailey\altaffilmark{12}, Travis S. Barman\altaffilmark{15}, Joanna Bulger\altaffilmark{16}, Sebastian Bruzzone\altaffilmark{17},  Jeffrey K. Chilcote\altaffilmark{18}, Tara Cotten\altaffilmark{19}, Rebekah I. Dawson\altaffilmark{20}, Robert J. De Rosa\altaffilmark{3}, Ren\'e Doyon\altaffilmark{2}, Thomas M. Esposito\altaffilmark{3}, Michael P. Fitzgerald\altaffilmark{21}, Katherine B. Follette\altaffilmark{12}, Stephen Goodsell\altaffilmark{22,23}, James R. Graham\altaffilmark{3}, Alexandra Z. Greenbaum\altaffilmark{24}, Pascale Hibon\altaffilmark{25}, Li-Wei Hung\altaffilmark{21}, Patrick Ingraham\altaffilmark{26}, Paul Kalas\altaffilmark{3}, David Lafreni\`ere\altaffilmark{2}, James E. Larkin\altaffilmark{21}, Bruce A. Macintosh\altaffilmark{12}, J\'{e}r\^{o}me Maire\altaffilmark{18}, Franck Marchis\altaffilmark{11}, Mark S. Marley\altaffilmark{27}, Brenda C. Matthews\altaffilmark{9,10}, Stanimir Metchev\altaffilmark{17,28}, Maxwell A. Millar-Blanchaer\altaffilmark{18,29}, Rebecca Oppenheimer\altaffilmark{8}, David W. Palmer\altaffilmark{14}, Jenny Patience\altaffilmark{30}, Marshall D. Perrin\altaffilmark{5}, Lisa A. Poyneer\altaffilmark{14}, Abhijith Rajan\altaffilmark{30}, Fredrik T. Rantakyr\"o\altaffilmark{23}, Dmitry Savransky\altaffilmark{31}, Adam C. Schneider\altaffilmark{32}, Anand Sivaramakrishnan\altaffilmark{5}, Inseok Song\altaffilmark{19}, Remi Soummer\altaffilmark{5}, Sandrine Thomas\altaffilmark{26}, J. Kent Wallace\altaffilmark{33}, Kimberly Ward-Duong\altaffilmark{30}, Sloane J. Wiktorowicz\altaffilmark{34}, Schuyler G. Wolff\altaffilmark{24}}
\altaffiltext{1}{Center for Astrophysics and Space Sciences,
  University of California, San Diego, La
  Jolla, CA 92093, USA; qkonopacky@ucsd.edu}
\altaffiltext{2}{Institut de Recherche sur les Exoplan\`etes,
  D\'epartement de physique, Universit\'e de Montr\'eal,
  Montr\'eal, QC H3C 3J7, Canada} 
\altaffiltext{3}{Astronomy Department,
  University of California, Berkeley; Berkeley, CA 94720, USA} 
\altaffiltext{4}{Univ. Grenoble Alpes/CNRS, IPAG, F-38000 Grenoble, France}
\altaffiltext{5}{Space Telescope Science Institute, Baltimore, MD 21218, USA}
\altaffiltext{6}{Department of Engineering Science and
  Physics, College of Staten Island, City University of New
  York, Staten Island, NY 10314, USA}
  \altaffiltext{7}{Physics Program, The Graduate Center, City University of New
  York, New York, NY 10016, USA} 
\altaffiltext{8}{Department of Astrophysics, American Museum of Natural History, New York, NY 10024, USA}
\altaffiltext{9}{National Research Council of Canada Herzberg,
  Victoria, BC, V9E 2E7, Canada}
\altaffiltext{10}{University of Victoria, Department of Physics
  and Astronomy, 3800 Finnerty Rd, Victoria, BC V8P 5C2, Canada}
\altaffiltext{11}{SETI Institute, Carl Sagan Center, Mountain
  View, CA 94043, USA} 
\altaffiltext{12}{Kavli Institute for Particle Astrophysics
  and Cosmology, Department of Physics, Stanford University,
  Stanford, CA, 94305, USA}
\altaffiltext{13}{Institute for Advanced Study, Princeton, NJ 08540, USA}
\altaffiltext{14}{Lawrence Livermore National Laboratory, 7000 East Ave, Livermore, CA, 94550, USA}
\altaffiltext{15}{Lunar and Planetary Lab, University of Arizona, Tucson, AZ 85721, USA}
\altaffiltext{16}{Subaru Telescope, NAOJ, 650 North A’ohoku Place, Hilo, HI 96720, USA}
\altaffiltext{17}{Department of Physics and Astronomy, Centre for Planetary Science and Exploration, The University of Western Ontario, London, ON N6A 3K7, Canada}
\altaffiltext{18}{Dunlap Institute for Astronomy \& Astrophysics, University of Toronto, 50 St. George St., Toronto, Ontario, Canada}
\altaffiltext{19}{Department of Physics and Astronomy, University of Georgia, Athens, GA 30602, USA}
\altaffiltext{20}{Center for Exoplanets and Habitable Worlds, 525 Davey Laboratory, The Pennsylvania State University, University Park, PA, 16802, USA}
\altaffiltext{21}{Department of Physics \& Astronomy, University of California, Los Angeles, CA 90095, USA}
\altaffiltext{22}{Department of Physics, Durham University, Stockton Road, Durham DH1, UK}
\altaffiltext{23}{Gemini Observatory, Casilla 603, La Serena, Chile}
\altaffiltext{24}{Department of Physics \& Astronomy, Johns Hopkins University, Baltimore MD 21218, USA}
\altaffiltext{25}{European Southern Observatory , Alonso de Cordova 3107, Vitacura, Santiago, Chile}
\altaffiltext{26}{Large Synoptic Survey Telescope, 950 N Cherry Ave, Tucson AZ, 85719, USA}
\altaffiltext{27}{Space Science Division, NASA Ames Research Center, Mail Stop 245-3, Moffett Field CA 94035, USA}
\altaffiltext{28}{Department of Physics \& Astronomy, Stony Brook University, Stony Brook, NY 11794-3800, USA}
\altaffiltext{29}{Department of Astronomy \& Astrophysics, University of Toronto, 50 St. George St., Toronto, Ontario, Canada}
\altaffiltext{30}{School of Earth and Space Exploration, Arizona State University, PO Box 871404, Tempe, AZ 85287, USA}
\altaffiltext{31}{Sibley School of Mechanical and Aerospace Engineering, Cornell University, Ithaca, NY 14853, USA}
\altaffiltext{32}{Department of Physics \& Astronomy, University of Toledo, 2801 W. Bancroft St., Toledo, OH 43606, USA}
\altaffiltext{33}{Jet Propulsion Laboratory, California Institute of Technology, 4800 Oak Grove Drive, Pasadena, CA 91109, USA}
\altaffiltext{34}{The Aerospace Corporation, 2310 E. El Segundo Blvd., El Segundo, CA 90245}


\begin{abstract}

We present the discovery of a brown dwarf companion to the
debris disk host star HR 2562.  This object, discovered with
the Gemini Planet Imager (GPI), has a projected separation of
20.3$\pm$0.3 au ($0.618\pm0.004\arcsec$) from the star.  With
the high astrometric precision afforded by GPI, we have
confirmed common proper motion of HR 2562B with the star with
only a month time baseline between observations to more than
$5\sigma$.  Spectral data in $J$, $H$, and $K$ bands show
morphological similarity to L/T transition objects.  We assign
a spectral type of L7$\pm$3 to HR 2562B, and derive a
luminosity of $\log$(L$_{\rm
  bol}$/L$_{\odot}$)=-4.62$\pm$0.12, corresponding to a mass
of  30$\pm$15 M$_{\rm Jup}$ from evolutionary models at an
estimated age of the system of 300--900 Myr.  Although the
uncertainty in the age of the host star is significant, the
spectra and photometry exhibit several indications of youth
for HR 2562B.  The source has a position angle consistent with
an orbit in the same plane as the debris disk recently
resolved with \textit{Herschel}.  Additionally, it appears to
be interior to the debris disk.  Though the extent of the
inner hole is currently too uncertain to place limits on the
mass of HR 2562B, future observations of the disk with higher
spatial resolution may be able to provide mass constraints.
This is the first brown dwarf-mass object found to reside in
the inner hole of a debris disk, offering the opportunity to
search for evidence of formation above the deuterium burning
limit in a circumstellar disk.    
\end{abstract}

\section{Introduction}

There is considerable interest in determining whether Jovian
planets on wide orbits represents a continuum that extends to
brown dwarf masses, or whether there is a strong cutoff in the
number of companions as a function of mass
\citep[e.g.,][]{kratter10}.  This relates to possible
formation pathways for substellar companions: either
companions form within a circumstellar disk and reach a mass
above the deuterium burning limit (e.g., \citealt{vorobyov13})
or via cloud fragmentation, as in binary systems with a high
mass ratio ($q$, \citealt{bate12}). Population statistics from
direct imaging provide essential observational parameters to
test formation history. From numerous surveys, only a handful
of imaged substellar companions are $<$100 au from their host
stars. In particular, this separation regime has shown a lack
of brown dwarfs with $q<0.1$ (the "brown-dwarf desert", e.g.,
\citealt[]{kraus08}) around stars with
M$>0.5$M$_\odot$. However, this parameter space has recently
begun to be populated by direct imaging
\citep[e.g.,][]{mawet15,hinkley15}.

Since the contrast of a substellar object
is more favorable for imaging with youth, direct imaging
surveys tend to target sources with evidence of a relatively
young age ($<300$ Myr).  The presence of a debris disk,
leftover of planet formation, is an insight for a
younger-than-field age since the dust luminosity is known to
decrease with time (see \citealt{wyatt08}). However, the dust
can persist to longer time if small planetesimals are
dynamically perturbed by orbiting companion, hence other youth
indicators are necessary to constrain the age of a star.  An
accurate estimate of the age of the star is necessary since
companion properties are often derived from evolutionary
models. Membership in nearby young associations
\citep[e.g.,][]{malo13} provides the tightest constraints on
the age of a star; otherwise, large variation in the derived
companion masses exists
\citep[e.g.,][]{kuzuhara13,fuhrmann15}.

Among the brown dwarf-mass companions that have been
discovered around stars with infrared excess, none have been
previously seen inside the inner hole of a resolved disk,
which offers the opportunity to study dynamical interactions.
Using the Gemini Planet Imager (GPI, \citealt{macintosh14}),
we report the discovery of a companion to the debris disk host
HR 2562. HR 2562B has a projected separation within the
"brown-dwarf desert", and within a possible cleared inner
hole.  We derive properties for this companion based on
spectra and colors, and discuss the potential roll it plays in
maintaining and shaping the debris disk.  

\section{HR 2562}

\begin{figure}

\begin{minipage}[b]{0.5\textwidth}
    \includegraphics[width=\textwidth]{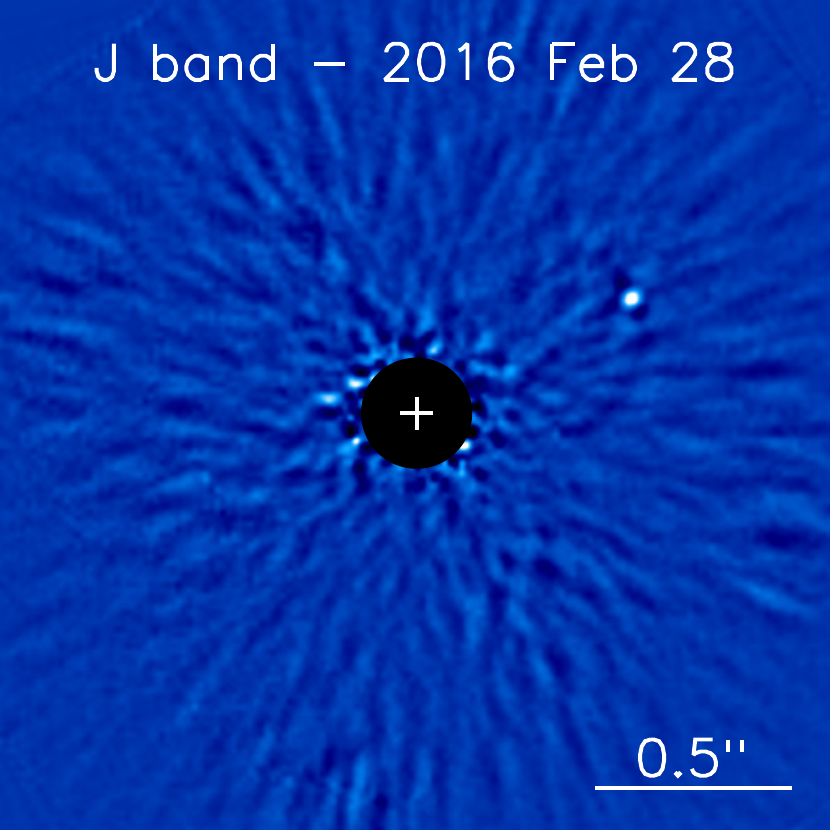}
  \end{minipage}
  \begin{minipage}[b]{0.5\textwidth}
    \includegraphics[width=\textwidth]{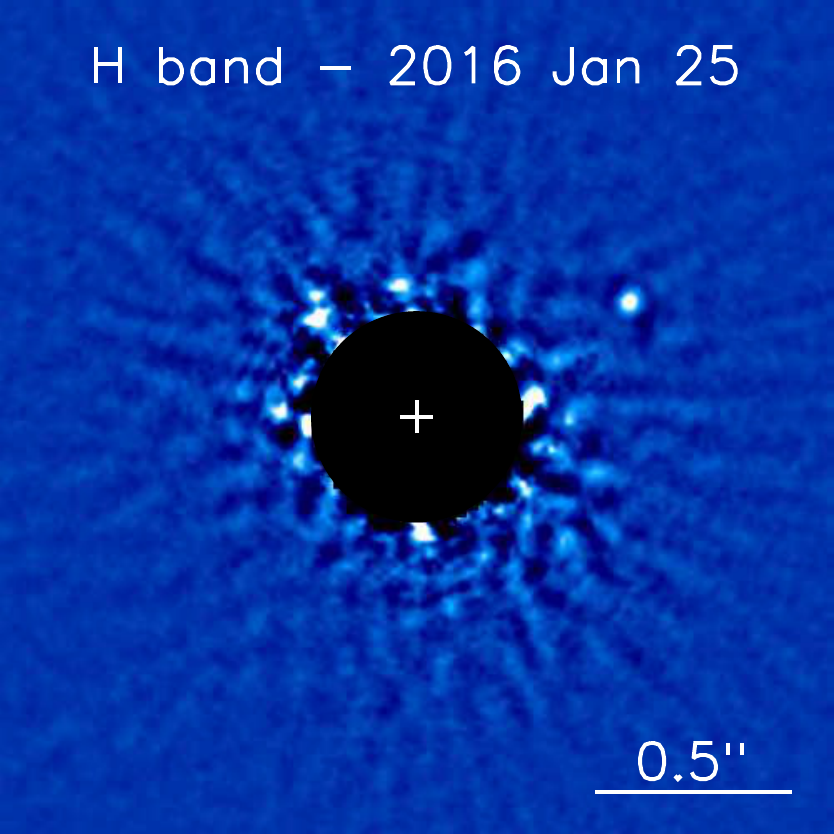}
      \end{minipage}
 \begin{minipage}[b]{0.5\textwidth}
    \includegraphics[width=\textwidth]{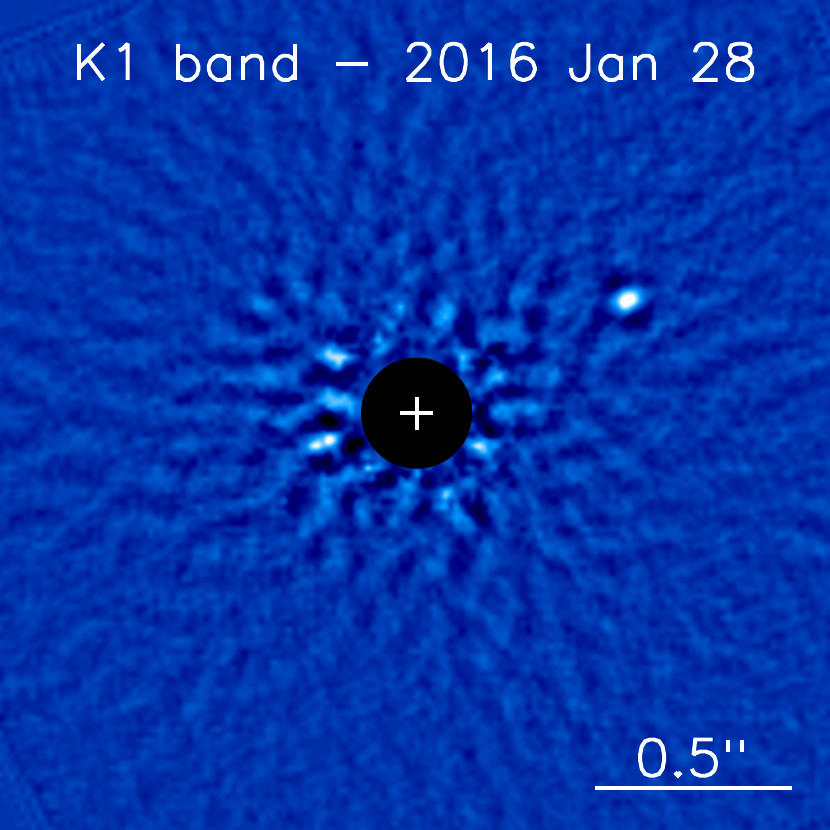}
      \end{minipage}
\begin{minipage}[b]{0.5\textwidth}
    \includegraphics[width=\textwidth]{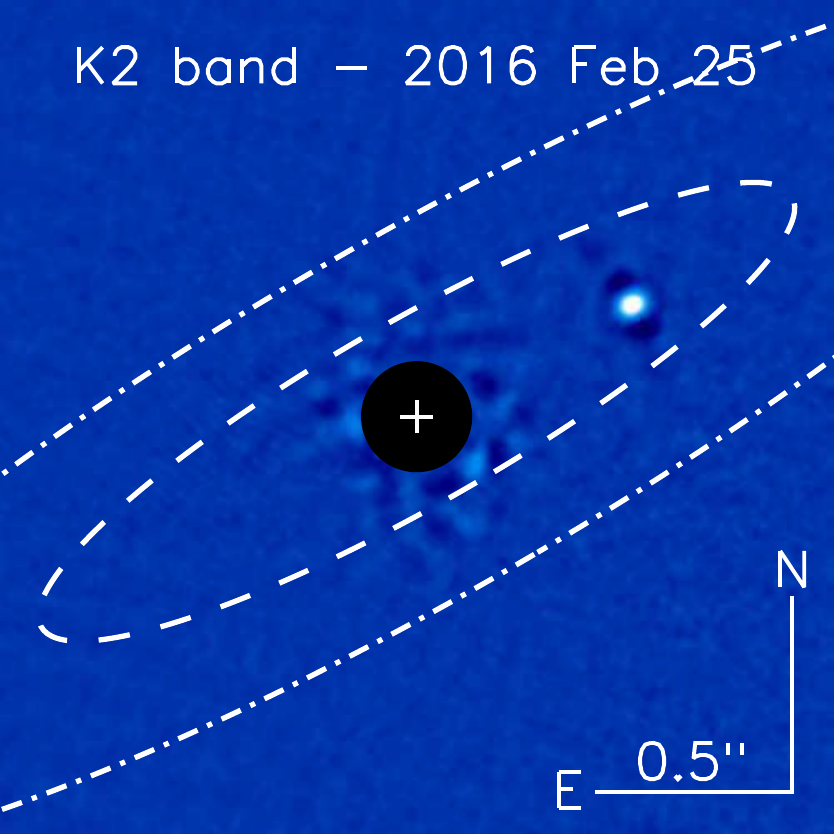}
      \end{minipage}
\caption{Collapsed datacubes showing HR 2562B in each of the
  four modes observed with GPI and reduced using KLIP. The
  $K2$ image is from February 2016 and demonstrates two
  possible solutions for the inner edge of the disk (38 and 75
  au with dashed and dotted-dashed lines respectively)
  assuming inclination of $78^\circ$ and position angle of
  $120^\circ$.}  
  \label{fig:images}
\end{figure}

HR 2562 is a F5V star with an estimated mass of 1.3
M$_{\odot}$ (\citealt{gray06,casagrande11}).  It has a
distance of 33.63$\pm$0.48 pc and proper motion of $\sim$ 110
mas/yr \citep{vanleeuwen07}.  It was identified as having a
debris disk with data from \textit{IRAS} and \textit{Spitzer}
by \citet{moor06}. \citet{gray06} identify the source as
active based on the Ca II H and K lines, while
\citet{torres00} identify it as an X-ray source.  Several
groups have also computed metallicity estimates, which range
from [M/H] = [-0.05,+0.08]
\citep{gray06,casagrande11,maldonado12}, with possible
evidence for peculiar individual abundances
\citep{casagrande11}.    

Currently there are several disparate age estimates for HR
2562 in the literature.  An estimate of 300$\pm$120 Myr was
made by \citet{asiain99}, who used a combination of space
motions and evolutionary model-derived ages to suggest that
the source is part of a nearby Local Association subgroup
called B3.  In their identification of the source using
\textit{IRAS}, \citet{rhee07} use space motions, a lithium
non-detection, and X-ray luminosity to give a rough age
estimate of $\sim$300 Myr. Conversely, analysis of data from
the Geneva-Copenhagen survey, in which metallicities and
temperatures are used to derive ages from models, gives an
estimated age of 0.9 - 1.6 Gyr (\citealt{casagrande11}).  In
their assessment of age based on Ca II H and K lines,
\citet{pace13} also derive an age of $\sim$900 
Myr.  Most recently, \citet{moor15} used photometric modeling
with atmosphere models to derive an age range of
300$^{+420}_{-180}$ Myr.  The BANYAN II group/field membership
estimation code \citep{gagne14} gives low probabilities for
the star to belong to any known young nearby group, but
suggests the star is younger than field stars, giving a wide
range of possible ages between $\sim$20 Myr to $\sim$1 Gyr.
While the age of the star remains uncertain, sufficient
evidence of moderate youth led to the inclusion of HR 2562 in
the sample for the Gemini Planet Imager Exoplanet Survey
(GPIES).  For the purposes of this paper, we adopt a nominal
age range of 300--900 Myr. 

\section{GPI Observations and Data Analysis}\label{data}

HR 2562 was observed in January 2016. GPIES observations are
taken in angular differential imaging (ADI,
\citealt{marois06}) $H$ band spectroscopic mode. A candidate
companion was identified in this initial data set.  Follow-up
observations were made within a month in the $K1$, $K2$, and
$J$ bands. Sky frames were also obtained right after the $K2$
sequence. Table 1 gives the log of these observations. Weather
conditions were median with DIMM seeing around $1\arcsec$ when
available. Another (longer) $K2$ sequence was acquired to
provide higher a signal-to-noise ratio for the companion and
was used for spectroscopy. All data were acquired with the
$H$-band apodizer providing a near-IR constant
star-to-satellite-spot\footnote{Satellite spots are
  diffraction spots created by a square grid placed in the
  pupil plane.} ratio of $9.23\pm0.06$ mag
\citep{perrin16}. Astrometric calibrator observations were
obtained in January and February 2016, and were analyzed as in
\citet{konopacky14}. These observations showed no change in
the IFS calibration as measured in previous GPIES
observations. Therefore, as in \citet{derosa15}, pixel scale
and position angle offset of $14.166\pm0.007$ mas/px and
$-0.10\pm0.13$ deg were used.   

Data were processed using the GPI data reduction pipeline
version $1.3.0$ (\citealt{perrin16}, and references therein)
to obtain calibrated $(x,y,\lambda)$ datacubes. Further
processing to suppress the star point spread function was
performed as described in \citet{macintosh15} and
\citet{derosa15}, using four independent pipelines and several
ADI algorithms: cADI \citep{marois06}, TLOCI \citep{marois14},
and KLIP/\textsc{pyKLIP} \citep{soummer12,wang15}. The
post-processed cubes were then combined to create broad-band
images, examples of which are shown in Figure
\ref{fig:images}. From each of these pipelines, positions and
contrast-per-slice and associated errors were extracted
following the forward modeled and minimization techniques
described in \citet{marois10,lagrange10}, and
\citet{pueyo16}. The final astrometric errors were combined in
quadrature from the errors on the measurements ($0.20-0.35$ px
depending on the dataset), astrometric calibration, and star
center ($0.05$ px). For the spectroscopy, the errors on the
measurements and on the star-to-spot ratio were similarly
combined. Final values for astrometry and contrasts for the
companion are derived by averaging the results of each
pipeline. Table 1 gives the astrometry and photometry derived
from each observation set. The spectra of the companion were
then obtained after normalization with a calibrated template
F5V spectrum from the Pickles library \citep{pickles98}, the
2MASS $JHK$ magnitudes of the host star and the GPI response
functions. Figure \ref{fig:spectrum} shows the final spectrum
from each bandpass. 

\begin{deluxetable}{lcccccccc} 
\tabletypesize{\scriptsize} 
\tablewidth{0pt} 
\tablecaption{Observations and Astrometry of HR 2562B} 
\label{tab:log}
\tablehead{ 
  \colhead{Date} & \colhead{Filter} & \colhead{$\lambda/\delta\lambda$} &
  \colhead{Total Int.} &
  \colhead{Field} &
  \colhead{$\rho$} & \colhead{$\theta$} & \colhead{Contrast} & \colhead{Absolute} \\
  \colhead{(UT)} & \colhead{} & \colhead{} & \colhead{Time (min)} &
  \colhead{Rot. (deg)} &
  \colhead{(mas)} & \colhead{(deg)} & \colhead{(mag)} & \colhead{Mag.} \\
}
\startdata 
2016 Jan 25 & $H$ & 45 & 37 & 20.2 & 619$\pm$3 & 297.56$\pm$0.35 & 11.7$\pm$0.1 &14.2$\pm$0.1 \\
2016 Jan 28 & $K1$ & 65 & 23 & 11.9 & 618$\pm$5 & 297.40$\pm$0.25 & 10.6$\pm$0.1 & 13.0$\pm$0.1 \\
2016 Jan 28 & $K2$ & 75 & 24 & 11.3 & 618$\pm$4 & 297.76$\pm$0.37 & 10.4$\pm$0.1 & 12.8$\pm$0.1 \\
2016 Feb 25 & $K2$ & 75 & 47 & 25.7 & 619$\pm$2 & 297.50$\pm$0.25 & 10.4$\pm$0.1 & 12.8$\pm$0.1 \\
2016 Feb 28 & $J$ & 35 & 54 & 26.6 & 620$\pm$3  & 297.90$\pm$0.25 & 12.6$\pm$0.1 & 15.3$\pm$0.1 \\
\enddata
\end{deluxetable}

\section{Properties of HR 2562B}

\subsection{Companionship}\label{cpm}

\begin{figure*}
\epsscale{0.95}
\plotone{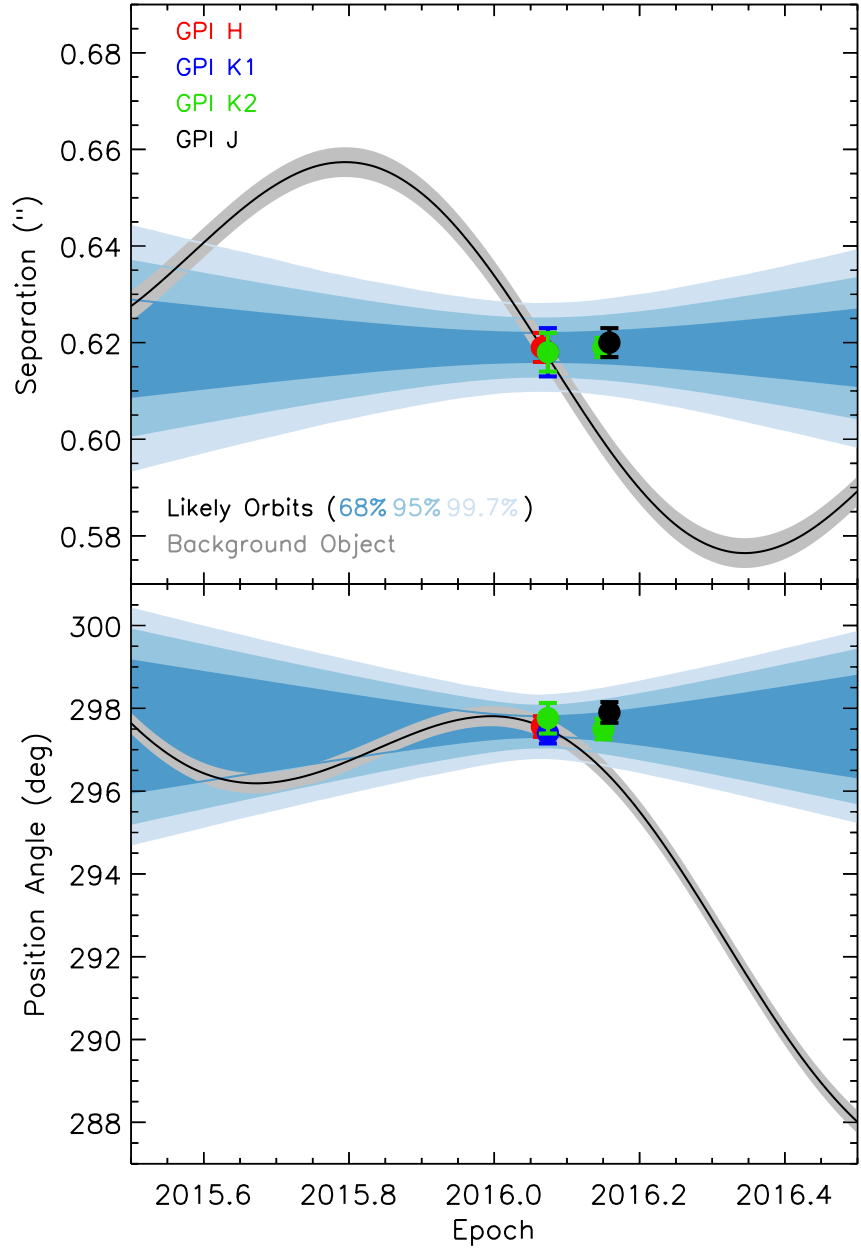}
\caption{Astrometric data points for HR 2562B as a function of
  epoch.  The gray lines show the path of a background object,
  while the blue shaded regions show the path of possible
  orbits for a bound companion.} 
\label{fig:astrometry}
\end{figure*}

At the distance of HR 2562, the projected separation of the
companion is 20.3$\pm$0.3 au ($0.618\pm0.003\arcsec$). We
assessed whether the candidate companion is comoving with the
star.  We compared all relative astrometric measurements with
the predictions of the location of an infinitely far
background object. The companion being an infinitely far
background object is ruled out to 5$\sigma$ \citep{nielsen13}.
Additionally, as in \citet{derosa15}, we estimate likely orbit
tracks for bound objects. The results of this assessment are
shown in Figure \ref{fig:astrometry}. The astrometry falls
clearly in the realm of bound orbits, with a preferred range
of semi-major axes of $\sim$15--42 au.    

Archival Hubble Space Telescope NICMOS data from 2007 was
found for HR 2562 (PI Rhee, ID 11157).  These data were
processed as part of the Archival Legacy Investigation of 
Circumstellar Debris Disks (ALICE, \citealt{choquet14}).
Although the contrast achieved in these images is insufficient
to detect the companion at its present location, we searched
for point sources at the location the companion would have
been if it was a background object ($\sim$1.3\arcsec).  No
source is detected at this location. 

As we demonstrate in Section \ref{spec}, the companion
photometry and spectroscopy are inconsistent with an
infinitely far background star. A more likely source of
contamination might instead be a non-stationary foreground or
slightly background L/T dwarf. Following the methodology in
\citet{macintosh15}, we determine the probability of finding a
L5-T5 type object in the GPI field of view by combining L and
T dwarf space densities\footnote{We assume a uniform space
  density of brown dwarfs, given that our observations could
  only detect L5-T5 objects at less than the scale height of
  the thin disk} and absolute magnitudes \citep{reyle10,
  pecaut13}. We calculate a false alarm probability of
7$\times10^{-7}$ within the GPI field of view for L5-T5
dwarfs. Since the companion was found after observing 203
targets, the final false alarm probability is
1.4$\times10^{-4}$. Therefore, the chance alignment of a
non-stationary, unbound brown dwarf is unlikely and the bound
status is more probable.

\subsection{Spectral Comparisons to HR2562B}\label{spec}

\begin{figure*}
\epsscale{1.0}
\plotone{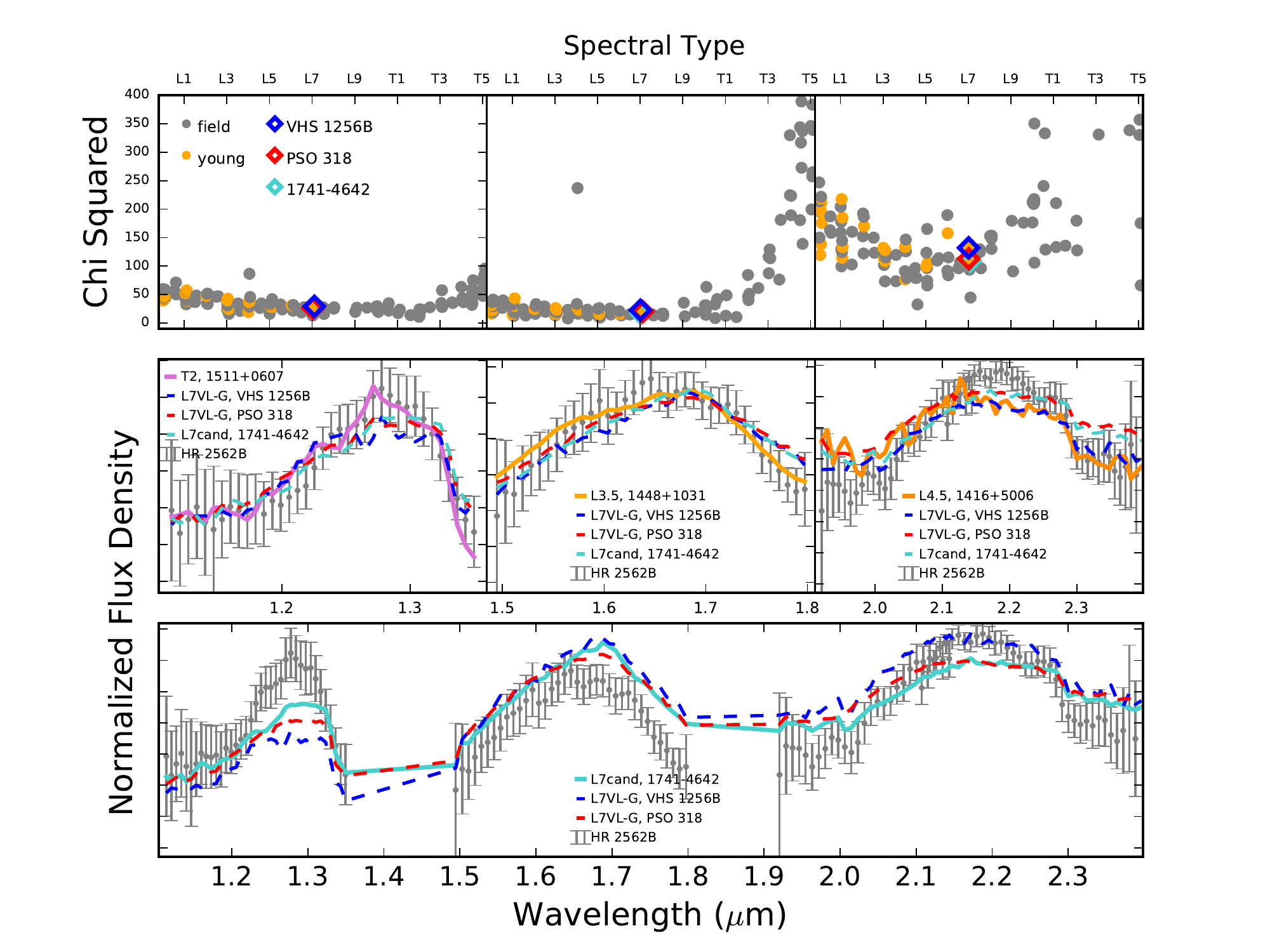}
\caption{The spectrum of HR 2562B (grey) shown with best
  fitting objects (solid lines) per band (middle row), and the
  best fit (solid aqua line, WISE J1741-4642) across all bands
  (bottom row).  Also shown are the young objects VHS 1256B
  and PSO 318 (dashed blue and red lines), which are both
  reasonable matches to the spectra. Corresponding $\chi^2$ as
  a function of spectral type is plotted for each band (top
  row). \label{fig:spectrum}} 
\end{figure*}

Photometric and spectral comparisons were made to assess the
likely properties of HR 2562B.  First, we used several
libraries of spectra and routines to determine the
best-matching spectral type for HR 2562B. We used the SPLAT
toolkit, which makes use of the SpeX prism library, to
determine which spectral types match HR 2562B
\citep{burgasser14}.  We find that the best-matching source
comparing all three spectral bands simultaneously is WISE
J174102.78-464225.5 (WISE 1741-4642), a recently discovered
peculiar L7$\pm$2 type brown dwarf with an estimated age of
10-100 Myr \citep{schneider14}.  Classifying the HR 2562B
using spectral standards in SPLAT returns a spectral type of
L7$\pm$0.5.

In a separate analysis, we used a $\chi^2$ goodness of fit
test with data from the SpeX Prism Library supplemented by
spectra from \citet{filippazzo15}.  The $\chi^2$ fits were
produced by normalizing the empirical spectra to HR 2562B with
a constant based on the flux and uncertainties of both
spectra, following \citet{cushing08}. The $\chi^2$ values were
then calculated between the empirical spectrum and HR 2562B
and assessed as a function of spectral type. In this analysis,
we considered all three bands simultaneously and separate fits
to each of the bands individually. We find that sources with
spectral types between L3.5 and T2 provide the best fits to
the data, depending on the band.  The best fit to the J-band
is a T2 type object, while L3.5 and L4.5 sources best match
$H$ and $K$, respectively.  The simultaneous fit to all bands
returns the same best fit as SPLAT, WISE 1741-4642.

In our analysis of $\chi^2$ as a function of spectral type, we
find that while earlier spectral types are preferred at H and
K, mid-to-late L types have nearly equivalent $\chi^2$.  We
also find that two other young L/T transition objects, VHS
J125601.92-125723.9B (VHS 1256B, \citealt{gauza15}) and PSO
J318.5338-22.8603 (PSO 318, \citealt{liu13}) are reasonable
matches to the spectra in individual bands.  When considering
the bands simultaneously, it is clear that the overall flux in
each wavelength is not perfectly matched by any other object,
including the best-fit WISE 1741-4642.  However, brown dwarfs
can have similar spectral features despite varying flux in
different wavelength bands (i.e., a range of colors), as
shown, for example, by \citet{leggett03} and
\citet{cruz16}. To investigate this, \citet{cruz16} created
band-normalized templates from optically-classified field
L0-L8 and L0$\gamma$-L4$\gamma$ objects and found that objects
with a range of $J-K$ colors as large as 0.60 can match
band-normalized templates with average $\chi^2>$0.9.  We
therefore adopt a spectral type of L7$\pm$3 for HR 2562B.
Figure \ref{fig:spectrum} shows the spectrum of HR 2562B,
along with the spectra of the best fitting objects in each
band and other similar young objects.

\begin{figure}

\begin{minipage}[b]{0.5\textwidth}
    \includegraphics[width=\textwidth]{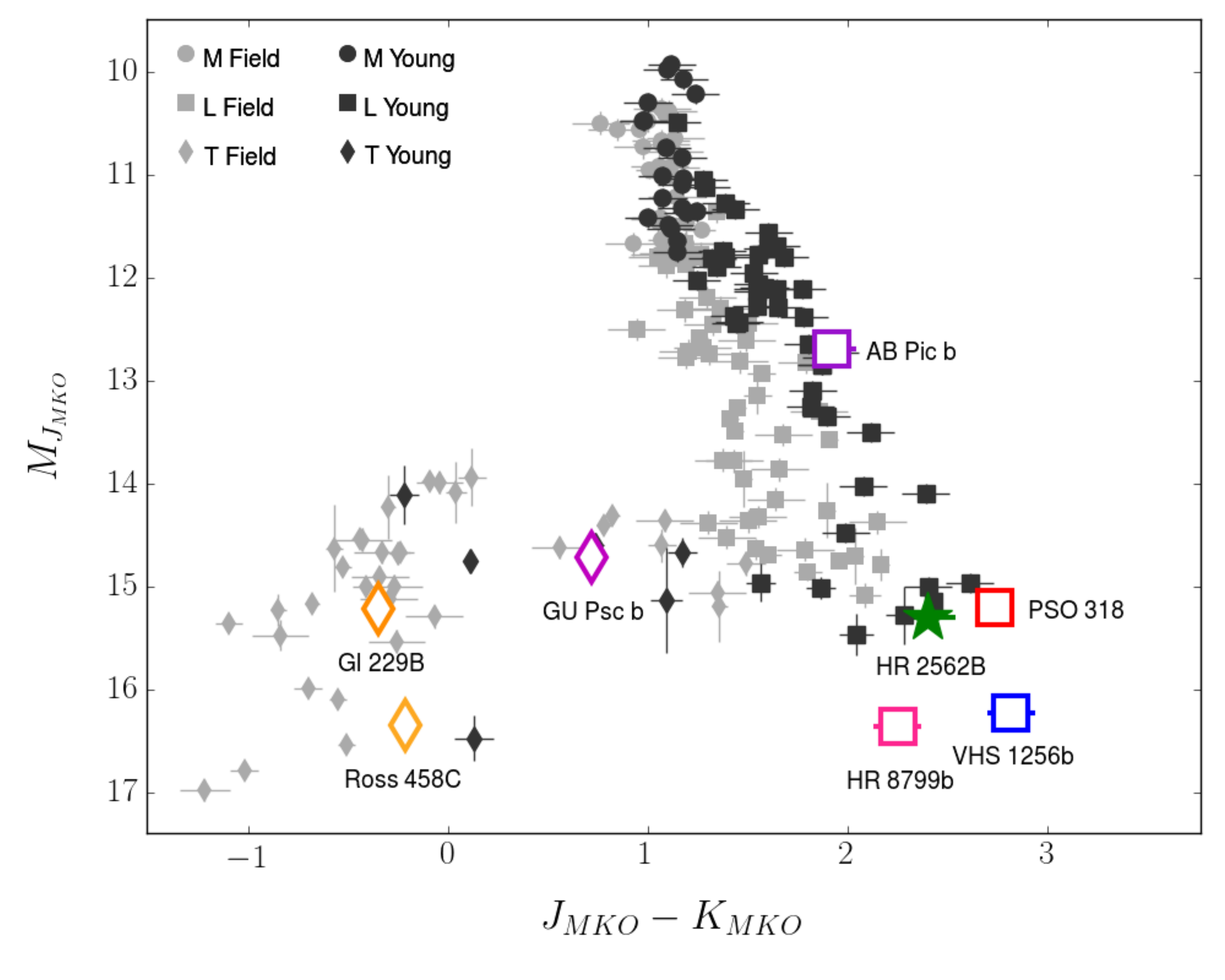}
  \end{minipage}
  \begin{minipage}[b]{0.5\textwidth}
    \includegraphics[width=\textwidth]{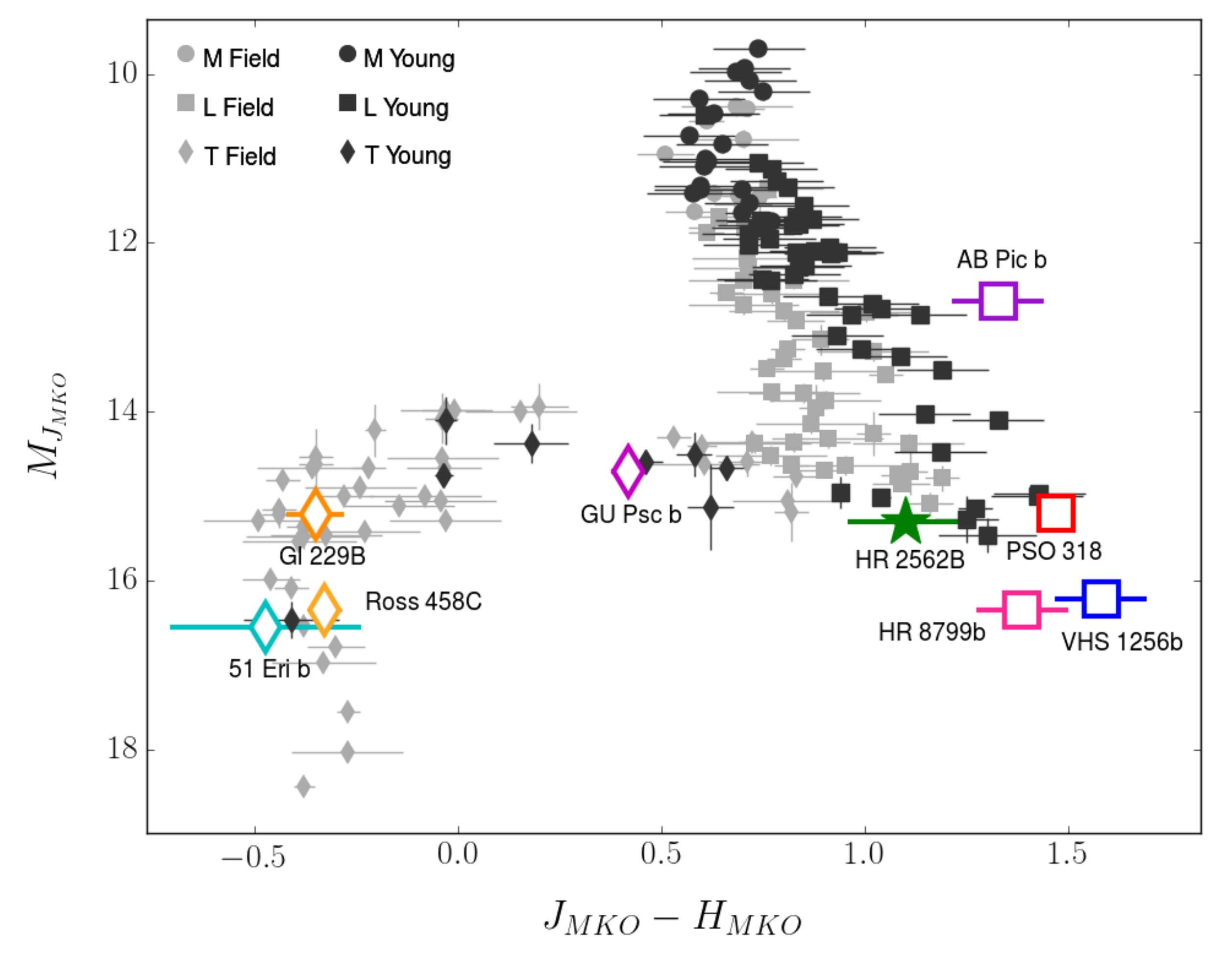}
      \end{minipage}
 \begin{minipage}[b]{0.5\textwidth}
    \includegraphics[width=\textwidth]{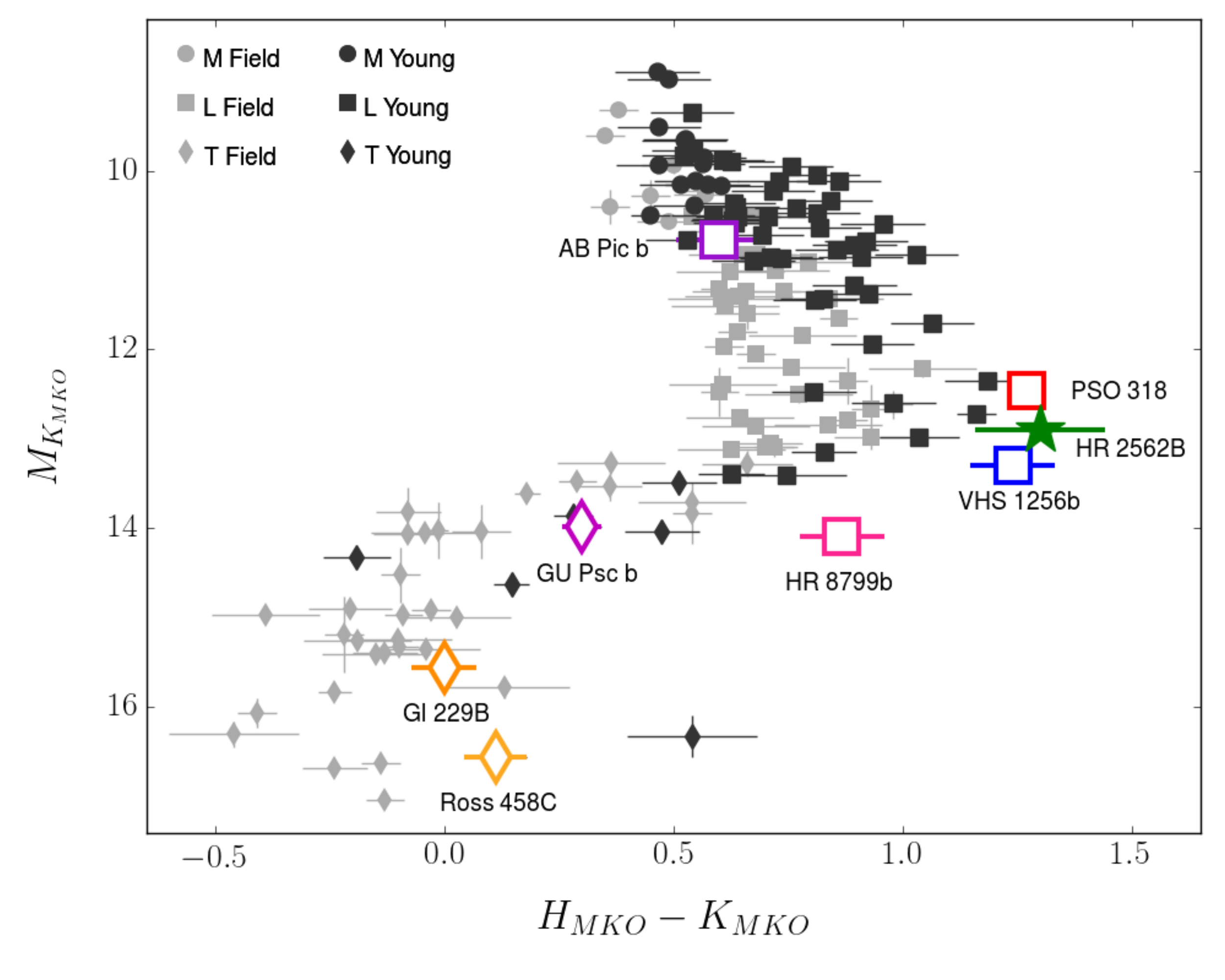}
      \end{minipage}
\begin{minipage}[b]{0.5\textwidth}
    \includegraphics[width=\textwidth]{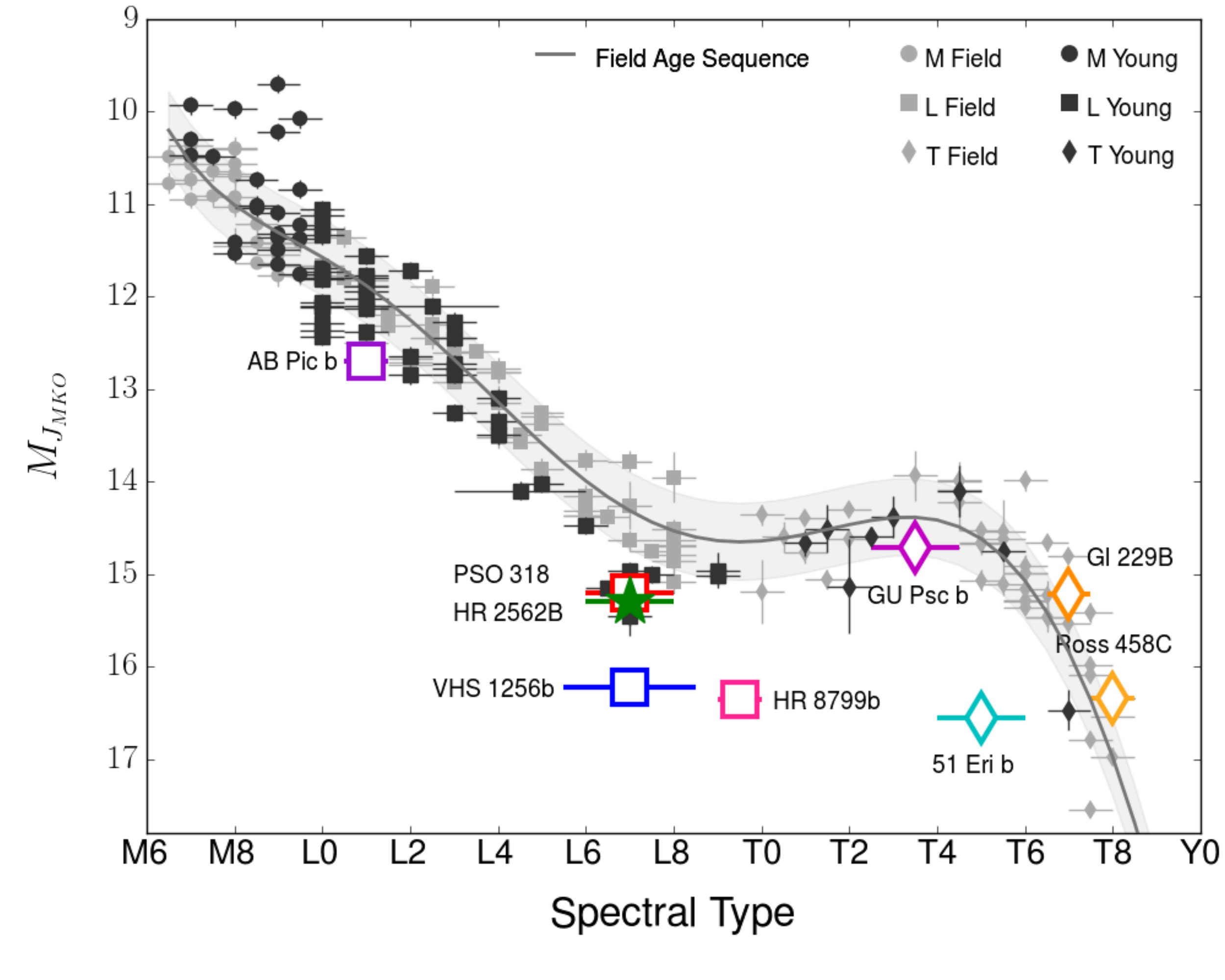}
      \end{minipage}
\caption{(Top Row and Bottom Left): Color-Magnitude diagrams
  ($JHK$) with the MLT sequence of 180 field objects (light
  grey symbols) from \citet{filippazzo15}. Young brown dwarfs
  or directly imaged companions are also shown (black
  symbols), and a few peculiar objects (specific symbols) from
  \citealt{Best:2015,faherty16}, Gagn\'e et al. (in
  prep.). (Bottom Right): The J band absolute magnitude as a
  function of spectral type. The grey line shows the field
  sequence with the $1\sigma$ spread marked by the grey shaded
  area following \citet{filippazzo15}.}  
  \label{fig:cmd}
\end{figure}

We then compare the colors of HR 2562B with other objects in
color magnitude diagrams (see Figure \ref{fig:cmd}). The
source clusters with the sequence of young, red, L-type
objects in the $M_J$ vs $J-K$ and $M_H$ vs $H-K$ diagrams.
Its $H-K$ is similar to VHS 1256b and PSO 318.  It is somewhat
bluer in $J-K$, though still near PSO 318 in absolute
magnitude at $M_J$.  Its consistency with young L type objects
is an additional suggestion of youth \citep[e.g.][]{gagné15}.   

We can also compare absolute photometry to other brown dwarfs
in both the field and young moving groups.  The variation of J
band magnitude with spectral type has been commonly used as a
method of distinguishing between field and potentially young
objects \citep[e.g,][]{faherty12}, with younger L-type objects
tending to be fainter at $J$ band than older objects of the
same spectral type.  This is thought to be a natural
consequence of changes in surface gravity with age impacting
atmospheric properties such as clouds (e.g.,
\citealt{marley12}).  Figure \ref{fig:cmd} shows the $J$ band
magnitude variation with spectral type. At a spectral type of
L7, HR 2562B clusters strongly with members of nearby moving
groups below the field population.

Following the methods of \citet{filippazzo15}, we constructed
a spectral energy distribution (SED) using the spectra and
photometry to determine an empirical $\log$(L$_{\rm
  bol}$/L$_{\odot}$) of $-$4.62$\pm$0.12.  Using this value
and an age range of 300--900 Myr, we then use the evolutionary
models from \citet{saumon08} to estimate the physical
properties of HR 2562B (solar metallicity, hybrid cloud).  We
find a mass of 30$\pm$15 M$_{\rm Jup}$, a radius of
1.11$\pm$0.11 R$_{Jup}$, a $\log(g)$ of 4.70$\pm$0.32, and a
temperature of 1200$\pm$100 K.  This temperature is somewhat
lower than field L7 type objects, but is consistent with
estimates for young objects of this spectral type.

\section{HR 2562B and the Debris Disk}\label{disk}

The identification of a brown dwarf in a system with a debris
disk presents interesting opportunities for constraining
system properties.  From \textit{Herschel} PACS, the disk is
marginally resolved. \citet{moor15} use this data to derive an
average dust radius of 112.1$\pm$8.4 au, with evidence for an
inner hole of radius between $\sim$18-70 au.  The average
outer radius is found to be 187 au.  Interestingly, they find
that the disk has a high inclination of 78.0$\pm$6.3$^{\circ}$
and a position angle on sky of 120.1$\pm$3.2$^{\circ}$.  Given
the separation of HR 2562B of $\sim$20 au and average position
angle of $\sim$298$^{\circ}$, the source appears to be
interior to the hole in the disk and coplanar with the disk to
within the uncertainties (see Figure \ref{fig:images}).  This
is the first example of a brown dwarf-mass companion within
the inner hole of a debris disk.

The existence of a companion interior to the disk provides the
exciting possibility to independently constrain the properties
of HR 2562 system.  \citet{moor15} discuss the possibility of
a self-stirring mechanism generating the resolved disk
material for a source as old as HR 2562A.  In this mechanism,
secular perturbations from a companion generate enhanced
collisions of smaller planetesimals, leading to the generation
of dust at wide separations \citep{mustill09}.  Using equation
6 in \citet{moor15} and assuming their nominal disk
parameters, we find that a mass of only 13M$_{\rm Jup}$ would
be required to generate collisions out to $\sim$187 au in 900
Myr assuming an eccentricity of 0.01.  If we use 30 M$_{\rm
  Jup}$ for HR 2562B, we derive a crossing time of $\sim$385
Myr, consistent with the lower end of our adopted age range.
While the uncertainties in the outer radius of the disk and
mass of the planet allow for crossing times $>$1 Gyr, the
nominal parameters are consistent with the scenario that HR
2562B is responsible for generating the observed disk.

A more complicated question is whether the inner hole can also
be used to place an upper limit on the mass of the companion.
Though \citet{moor15} derive an average inner radius of 38
$\pm$ 20 au from \textit{Herschel} images, the uncertainty on
this value is fairly large. \citet{moor15} also performed a
separate SED fit to available photometry and derived an inner
radius of 64 $\pm$ 6 au.  Since these two values are not
consistent, we performed a quick analysis in which we
simultaneously fit the SED and the \textit{Herschel} PACS
image using MCFOST \citep{pinte06}.  We used the geometric
parameters from \citet{moor15}, a flat surface density
profile, and a minimum grain size of $\sim$micron.  We find
that the SED and image together are best reproduced using an
inner radius of $\sim$75 au.  This radius is consistent with
upper end of the uncertainty range and SED fit of
\citet{moor15}.  For completeness, we use both an inner radius
of 38 au and our derived value of 75 au to roughly determine
whether mass constraints are possible.

Using the dynamical stability criterion proposed by
\citet{petrovich15}, we estimate the mass of the HR 2562B
assuming it is responsible for clearing the inner hole and
that it has an eccentricity of zero.   We find that if the
inner hole is 38 au, the upper limit on the mass of the
companion is $\sim$20 M$_{\rm Jup}$.   The difference between
an age of 300 Myr and 900 Myr is negligible given our other
assumptions.  If instead we use an inner radius of 75 au, the
upper limit on the mass is $\sim$0.24 M$_{\odot}$, well beyond
the highest estimates from evolutionary models.  If the inner
radius is indeed $>$75 au, it might suggest an elevated
eccentricity for HR 2562B.  Future observations that constrain
the orbital parameters of the companion and high resolution
images of the disk will offer insight into the potential
history of interaction between the bodies in this system and
provide meaningful mass limits. 

\section{Conclusions}

The HR 2562 system offers a relatively rare opportunity to
probe the direct dynamical interaction of a substellar object
with a Kuiper Belt analog.  The overall system architecture
may provide interesting clues to the formation of the
companion.  With a mass ratio of $q=0.02\pm0.01$, HR 2562B is
a new object in the growing list of substellar companions
within $30$ au \citep[e.g.,][]{mawet15,hinkley15}. These are
excellent candidates for formation via disk instability, which
has been shown to naturally produce objects as massive as 42
M$_{\rm Jup}$ at separations $\gtrsim 70-100$ au
\citep[e.g.,][]{rafikov05,kratter10}. Several challenges to
this picture remain, however, particularly the proximity of
observed objects to their host stars. At such close
separations the fast cooling needed for the disk fragmentation
into bound objects becomes difficult to realize, precluding
in-situ formation of these brown dwarfs by the gravitational
instability \citep[e.g.,][]{rafikov05}.  However, it is
plausible that the relatively massive HR 2562B formed beyond
50-70 au and migrated inward to its current location
\citep[e.g.,][]{vorobyov13}. Constraining the true mass and
orbit of the companion is essential to determining its
possible origin, which could offer evidence of “planet”
formation above the deuterium burning limit. 

\acknowledgements
The authors thank Richard Gray for his clarifying points on
spectral classification.  We also thank Adam Burgasser and
Daniella Bardalez-Gagliuffi for helpful discussions.  We also
thank an an anonymous referee whose comments improved this
manuscript.  This research has benefited from the SpeX Prism
Library and SpeX Prism Library Analysis Toolkit, maintained by
Adam Burgasser at http://www.browndwarfs.org/spexprism, from
the BANYAN II web tool at
http://www.astro.umontreal.ca/~gagne/banyanII.php?targetname=HR+2562\&resolve=Resolve,
and from the SIMBAD database, operated at CDS, Strasbourg,
France. Based on observations obtained at the Gemini
Observatory, which is operated by the Association of
Universities for Research in Astronomy, Inc., under a
cooperative agreement with the National Science Foundation
(NSF) on behalf of the Gemini partnership: the NSF (United
States), the National Research Council (Canada), CONICYT
(Chile), the australian Research Council (australia),
Minist\'{e}rio da Ci\^{e}ncia, Tecnologia e Inova\c{c}\~{a}o
(Brazil) and Ministerio de Ciencia, Tecnolog\'{i}a e
Innovaci\'{o}n Productiva (Argentina). J.R., R.D. and
D.L. acknowledge support from the Fonds de Recherche du
Qu\'{e}bec. Supported by NSF grants AST-1518332 (R.J.D.R.,
J.R.G., J.J.W., T.M.E., P.K.), AST-1411868 (B.M., A.R.,
K.W.D.), AST-141378 (G.D.), AST-1211568 (P.A.G., E.L.R.),
DGE-1232825 (A.Z.G.), and AST-1313132 (J.F.C.,
E.L.R.). Supported by NASA grants NNX15AD95G/NEXSS and
NNX15AC89G (R.J.D.R., J.R.G., P.K., J.J.W., T.M.E.), and
NNX14AJ80G (E.L.N., S.C.B., B.M., F.M., M.P.). Portions of
this work were performed under the auspices of the
U.S. Department of Energy by Lawrence Livermore National
Laboratory under Contract DE-AC52-07NA27344 (S.M.A., L.P.,
D.P.).

\end{document}